# A Methodology for Questionnaire Analysis: Insights through Cluster Analysis of an Investor Competition Data

Carlos Henrique Q. Forster[1],  Paulo André Lima de Castro[1]*  and  Andrei Ramalho [2]

*Aeronautics Institute of Technology, Brazil[1]*
*XP Investimentos, Brazil[2]*
* Corresponding author's Email: pauloac@ita.br

**Abstract:** In this paper, we propose a methodology for the analysis of questionnaire data along with its application on discovering insights from investor data motivated by a day trading competition. The questionnaire includes categorical questions, which are reduced to binary questions, "yes" or "no". The methodology reduces dimensionality by grouping questions and participants with similar responses using clustering analysis. Rule discovery was performed by using a conversion rate metric. Innovative visual representations were proposed to validate the cluster analysis and the relation discovery between questions. When crossing with financial data, additional insights were revealed related to the recognized clusters.

**Keywords:** Exploratory data analysis, cluster analysis, dimensionality reduction, rule set mining, icon-based visualization, pie charts, binary data, Bernoulli distributed data, questionnaire analysis, day trading, future markets.

## 1. Introduction

In today's rapid evolving digital world with vast amounts of data being generated daily. Efficient data analysis is highly relevant. In the finance field, understanding investor's behaviour may help stock brokers and other financial service providers to offer more suitable produtcts [1]. Furthermore, they may provide intelligent systems to support their investor decision process [2].

There are many possible sources of data, that can be used for analyze behaviour, for instance data from social media [3]. However, in order to better understand investor behaviour, it can be really helpful to analyze data from real operation and direct answers from real investors. ,A great opportunity to acquire such information took place in May 2022 during an investor competition regarding day trade operations of the clients of XP, one of the main brokers in Brazil. The clients that subscribed to the competition were entitled to prizes according to their measured performance in day trade investing. Along with this, it was possible to run a questionnaire in order to understand better the background and objectives of the competitors.

In this paper, we describe and apply a methodology to extract insights from the questionnaire data. The exploratory analysis consists on many technique applied to data corresponding to the answers to seven question groups, where each answer is binary: either selected or not by the respondent.

A key to this analysis is the detection of common behavior among the participants. A dimension reduction will reveal question selections that co-occur in the participant's answers. It also leads to a simplification of further analysis methods. Cluster analysis allows the definition of profiles that are able to represent groups of participants with similar characteristic behaviors. It also allows us to segment the whole set into a fixed number of profiles for individualized analysis. Finally, data reduction can be used to improve the number of questions by selecting the most informative and representative, easing analyzes that were impacted by the high dimensionality of the data.

For our analyzes we also propose innovative graphical representations. The GrapeShape charts allow the iconic visualization of the clusters, revealing their diversity level and internal variance by using a metaphor of a colored bunch of grapes. The HalfPie charts allow the observation of how one binary variable affects the proportion of the other and embeds the confidence interval of a proportion hypothesis test.

By crossing financial information with the questionnaire, we show that the investor clusters are related to the rank of financial returns. We also devise and apply a relation mining method in order to extract both expected and unexpected insights relating pairs of questions.

## 2. Related work

It is usual to employ Principal Component Analysis to find dimensions defined by a set of weights multiplying each response value 0 or 1. PCA [5] has the advantage of finding orthogonal axes of maximal variance and also it is easy to find the appropriate number of components. Each respondent will be characterized by a small vector of real numbers that weight each component vector to reconstruct one's response. Thus, interpretation would happen in this real-number space, which is a different method from what we proposed, that relies on a set of representative questions with Bernoulli distributed binary [6] values instead.

There are theoretical alternatives to clustering that relies on the likelihood formula of the Benoulli Mixture Model (BMM) [6]. However, k-means clustering algorithm [5] already does a good job of finding a grouping with good likelihood, while the BMM method may be hard to converge to expected results. It usually adopts k-means as an initial step, it has problems with proportions 0% and 100% that generate null probabilities. Rather than likelihood, we desire to improve intraclass variance, which is already provided by the k-means approach [10].

Concerning visualization of high dimensional data, techniques based on Chernoff faces [6] are applicable, but it would plot a median face representation of a cluster and the limits of each feature is not known from the plot. Our method can clearly display the profiles of each cluster allowing visual comparison to verify their diversity and also has the capability to display the variance of the cluster and its features. The color scale allows us to know how close each feature is to the limit (which has a 100% certainty) or close to the central color of the scale (green grapes, for features that have maximum variance). Other techniques such as parallel-category plots will also present their own difficulties to provide the information we need.

For rule mining between two variables, many confusion-matrix-based metrics can be used instead ours. For instace, we can mention chi-squared, McNemar, or just plain accuracy. Our proposed conversion rate metric offers a more interpretable view when used along the HalfPie charts, including hypothesis test.

## 3. Questionnaire description

The data analyzed in this study were provided by Corretora Clear [7] from the XP group [8]. Before being made available, such data was anonymized in order to completely prevent the identification of people and guarantee their privacy.

The analyzed data encompass three large groups of data. The first consists of customer information totaling 21,342 customers. The second group consists of responses voluntarily provided by such clients to a seven-question multiple-choice questionnaire proposed to clients. There were a total of 3067 customers who responded to the questionnaire. Finally, the third group of data consists of data on operations carried out by customers in day trade operations (started and finished on the same day). There are more than 44 million operations in this group in the period from 29/Dec/2021 to 31/Aug/2022 (approximately 8 months).

Operations data was aggregated by customer and stratified using customer related data and responses available in the questionnaire, in order to bring insights and identify hidden patterns in the data.

Regarding the questionnaire (Table 1), each question group contains a number of response items, which may or may not be checked by the respondent. For example, question group 1 has the following possible answers: Q1A, Q1B, Q1C etc, and may have multiple selected. Thus, each response contains a "yes" or "no" value associated with the customer who responded. Question group 4 is the only one where it is forced to have only one true answer, but we kept it as multiple dummy variables to facilitate the cluster analysis. Responses are assumed as Bernoulli distributed.

The questionnaire was optional and the respondents were voluntary and self-appointed. The lack of noticeable inconsistencies and the rediscovery of already expected relations indicate that respondents took seriously the task. The population are investors in the day trade modality that were subscribed to the competition.

## 4. Methods and Analysis

The exploratory analysis methodology we propose considers a questionnaire with binary responses, assumed Bernoulli distributed. A Bernoulli variable X with parameter p (proportion of positive responses) has a theoretical variance given by Eq.1:

(1)

where T is the expected number of "yes" and F is the expected number of "no" responses, being N the total number of responses. The variables are ordered according to their variances because a response to a high variance question is more informative than to a low one. Maximum variance is 0.25 when p=0.5, which is the case of maximum uncertainty.

Table 1. Questionnaire part 1.

| 1: How did you discover day trading? |
|---|
| Q 1 A - Friends & Family |
| Q 1 B - Investment Advisory |
| Q 1 C - Brokers |
| Q 1 D - Influencer / Youtube |
| Q 1 E - Other financial institutions (eg Banks) |
| Q 1 F - Advertising or Publicity |
| Q 1 G - News and Finance Website |
| **2: Which channels do you use to keep up to date?** |
| Q 2 A - Friends |
| Q 2 B - Investment Advisory |
| Q 2 C - Discord |
| Q 2 D - Facebook |
| Q 2 E - Instagram |
| Q 2 F - I don't follow news |
| Q 2 G - Podcast |
| Q 2 H - News & Finance Sites |
| Q 2 I - Telegram |
| Q 2 J - Twitter |
| Q 2 K - WhatsApp |
| Q 2 L - Youtube |
| **3: What are your main motivations for entering the modality?** |
| Q 3 A - Seek High Financial Return |
| Q 3 B - Compete with friends |
| Q 3 C - Learn more about the financial market |
| Q 3 D - Develop a career in the financial market |
| Q 3 E - Generate monthly income |
| Q 3 F - Interacting socially or belonging to a group |
| Q 3 G - Having a Hobby or Pastime |
| Q 3 H - Having an intellectual challenge |
| Q 3 I - Working autonomously and flexibly |
| **4: How long have you had experience with day trading?** |
| Q 4 A - Up to 1 month |
| Q 4 B - From 1 to 3 months |
| Q 4 C - From 12 to 24 months |
| Q 4 D - From 3 to 6 months |
| Q 4 E - From 6 to 12 months |
| Q 4 F - More than 24 months |

Table 2. Questionnaire part 2.

| 5: What is your level of investment knowledge? |
|---|
| Q 5 A - Watched some videos and searched on the subject on the internet |
| Q 5 B - Read books on the subject |
| Q 5 C - Completed more than 40 hours in a complete course or several smaller ones |
| Q 5 D - Completed less than 40 hours in one or more short courses |
| Q 5 E - Do you have an undergraduate or graduate degree in the area (finance, economics, etc.) |
| Q 5 F - Do you work in the financial market |
| **6: What investment products have you studied?** |
| Q 6 A - Cryptocurrencies |
| Q 6 B - Multimarket Fund |
| Q 6 C - Equity Funds |
| Q 6 D - Non-Financial Investments (Real Estate, Land, Jewelry, Art, etc.) |
| Q 6 E - BM&F Market (Mini index, dollar, commodities, etc.) |
| Q 6 F - Bovespa Market (stocks) |
| Q 6 G - International Market / Exchange |
| Q 6 H - Structured Operations |
| Q 6 I - Fixed Income |
| Q 6 J - Treasury Direct |
| **7: After you start day trading, you...** |
| Q 7 A - Tracks but does not actively interact with other day traders |
| Q 7 B - Does not follow or interact with other day traders |
| Q 7 C - Began to integrate day trading digital communities (social networks, groups) |
| Q 7 D - Started interacting with other day traders |
| Q 7 E - Started publishing content related to day trading on its social networks |

In order to perform an exploratory analysis of the questionnaire responses, the main technique is cluster analysis. Cluster analysis is applied with two objectives. The first is clustering the questions (columns) in order to reduce dimension. The second is clustering the responses (rows) in order to group the respondents identifying common behavior.

### 3.1 Clustering of questions.

If there is a large number of questions, many of which are correlated and others of little relevance, this space is reduced by selecting a subset of questions that have the following characteristics: to be representative of the information in the questionnaire responses as a whole, to be diversified, to be semantically important for raising hypotheses and results of interest.

For the selection of a set of questions, a cluster analysis is performed and within each group, the most representative questions are selected through the subjective weighting of several criteria, usually one per cluster. Among the criteria are: subjective criterion of the purpose of the analysis, the variables of maximum uncertainty, the variables highlighted by the monothetic grouping. Maximum uncertainty variables are interesting because, once their value is known, it contributes more to the global reduction of uncertainty.

Monothetic clustering defines a variable that will be used to divide the set of respondents into two subgroups. Recurrently, each subgroup formed is divided in turn by another variable chosen for each one. This is done up to a given defined depth. The idea of using the monothetic clustering is to minimize the variance of the groups after being divided and thus identify which were the most descriptive fields in the sense of minimizing the variance. For this, an exhaustive search was used until the division into 16 groups. Each element goes through three decisions before being assigned to a group. The search defined which questionnaire responses corresponded to the 15 bifurcations. The monothetic clustering produces the categorization in Fig. 1.

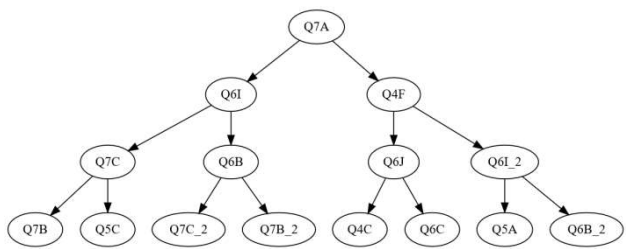

Figure 1. Monothetic grouping decision tree highlighting the responses that most reduce uncertainties (sum of intraclass variances) in the formation of 16 groups of respondents

We chose the Euclidean distance as a means of comparison, where a response corresponds to a vector of numbers 0 (for "no") and 1 (for "yes") for each respondent in the questionnaire. This turns out to be the square root of the number of respondents who assigned different values to the answer.

Thus, if two responses are in the same group, then it is expected that a respondent will more often associate the same value to these two responses (either "yes" or "no"). We consider that the questionnaire dataset is a matrix in which each line corresponds to a respondent and each column corresponds to a response. Thus, each cell of the matrix must contain the number 1 if the answer value for that respondent is "yes", or 0 otherwise.

The K-means algorithm was first applied to the questionnaire database to perform this column selection, simplifying the analysis when appropriate. K-means was set to initialize by the K-means++ technique, running 2000 times with random starts, returning the solution that minimizes the so-called "inertia" which is the sum of the squared distances of each element to the nearest centroid and which therefore corresponds to the group of that element.

From the dimensionality reduction, 10 groups were formed as depicted in Fig. 2, considering a representative response in bold for each group.

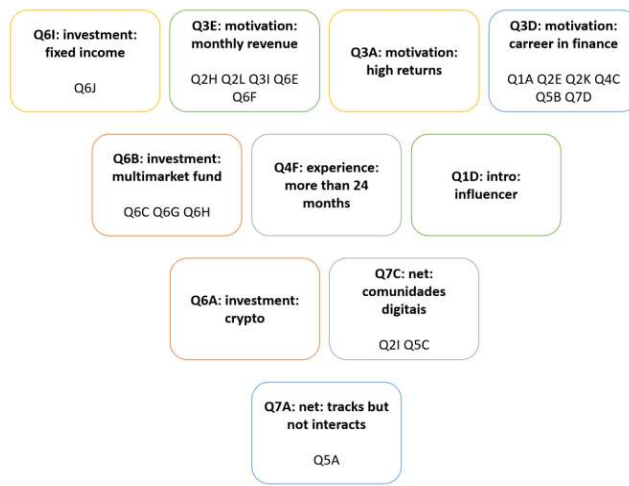

Figure 2. Grouping of questionnaire options by co-occurrence. One representative question was chosen for each group and highlighted.

Question Q6A crypto was analyzed separately from its cluster (Q6B, Q6C, Q6G, Q6H) as it is a phenomenon that we wanted to observe in more detail. The cluster formed by the other questions (Q1B, Q1C, Q1E, Q1F, Q1G, Q2A, Q2B, Q2C, Q2D, Q2F, Q2G, Q2J, Q3B, Q3C, Q3F, Q3G, Q3H, Q4A, Q4B, Q4D, Q4E, Q5D, Q5E, Q5F, Q6D, Q7B, Q7E) are questions with low variance and few positive answers. Such group is exempted from the analysis because it seems to be a grouping of items of lesser relevance and informative content (due to lesser variance).

See that the elements of the decision tree confirm many of the items discovered from the choices made. In particular, the choice of Q6B Multimarket as representative of group 0 was based on this method to the detriment of Q6C with higher variance. As well as the Q7C Communities preference over the Q5C Course of more than 40 hr, but which continues to be a very relevant variable to be studied.

Table 3. Choice of grouping method and number of clusters using internal metrics. In bold, the results that lead to the decision.

|  | Silhouette Score | Calinski Harabasz | Davies Bouldin |
|---|---|---|---|
| KMeans 2 | 0.125 | 433.939 | 2.644 |
| KMeans 3 | 0.116 | 372.893 | 2.399 |
| KMeans 4 | 0.116 | 329.651 | 2.312 |
| KMeans 5 | 0.109 | 296.316 | 2.208 |
| KMeans 6 | 0.116 | 273.165 | 2.361 |
| KMeans 7 | 0.110 | 251.889 | 2.207 |
| KMeans 8 | 0.115 | 236.217 | 2.179 |
| KMeans 9 | 0.108 | 224.605 | 2.090 |
| KMeans 10 | 0.115 | 211.703 | 2.120 |
| KMeans 11 | 0.116 | 199.698 | 2.043 |
| KMeans 12 | 0.115 | 194.447 | 2.023 |
| KMeans 13 | 0.118 | 186.715 | 2.068 |
| KMeans 14 | **0.127** | 184.272 | **2.000** |
| KMeans 15 | 0.126 | 175.417 | **1.931** |
| Agglomerative | 0.063 | 144.106 | 2.503 |
| Agglom.L1 | 0.054 | 123.344 | 2.233 |
| Spectral | 0.107 | 191.651 | 2.132 |

### 3.2 Clustering of respondents.

The grouping of instances is the process that effectively generates customer profiles based on questionnaire responses. The expectation is that each customer profile will show similar behavior, but that the profiles will be different. The combination of profiles covers all respondents.

A difficult question is the definition of the number of clusters (profiles). There is no single method for all situations and there is often no consensus on the best method. Thus, to avoid the subjectivity of the choice as much as possible, several grouping algorithms were applied and varying the number of clusters. Some metrics were used to select the best configuration.

The K-means algorithm is considered the most popular. The algorithm relies on defining distance (usually Euclidean) and calculating a centroid from a set of points. But even so, it can be applied to binary variables, considering for example values 0 and 1. In the case of categorical variables, it is recommended to convert them into binary variables (one-hot representation). K-means generates cluster centroids using the mean. This coincides with the estimation of the parameter of the Bernoulli distribution, that is, the proportion of positive responses.

The natural distance between vectors of binary variables would be the Hamming distance. However, some methods generate centroids with intermediate values from 0 to 1, which are very representative of the proportions of each variable. Thus, the L1 distance (Manhattan) can be used to compare the instances with these centroids, as it generalizes the Hamming distance.

The L2 (Euclidean) distance is interesting because it allows, through its summation, to calculate a very good approximation of the variance of each grouping based on centroid (its inertia). And it is used in K-means.

A problem with K-means is the instability of the solution. Because it relies on random numbers to draw, it can produce a different solution each round. Thus, we chose to set the seed of the random number generator and run the algorithm a large number of times, selecting the solution that obtains the minimum "inertia", which in terms of k-means means the sum of squares of the distances of each element to the their respective centroid. This is a good estimate of the intraclass variance, which is the real purpose of clustering, to obtain classes that have their elements distributed with small variance.

We compare the K-means applied to the original variables with the K-means applied to the variables selected for dimensionality reduction and we show that there is a great advantage in dimensionality reduction and that therefore it was adopted to generate the final grouping.

Other methods tested, but without further elaboration, were Ward's hierarchical agglomerative method and the spectral method.

Of the clusters evaluation metrics, there are external metrics and internal metrics. However, external metrics depend on a reference (ground-truth) with true classes, which we do not have in this analysis. This is because we don't know the profiles and we want to discover them from the data.

Internal metrics are those that observe the structure obtained from the classes, and we use the following three metrics: Davies-Bouldin (varies from 0 to infinity and the smaller the better), Calisnki-Harabasz [9] (it is a number that the larger the better), Silhouette (ranges from -1 to 1, the higher the better).

Table 4. Description and stereotype of the 14 customer profiles found by the algorithms used.

| Class | Description | Stereotype |
|---|---|---|
| 0 | Goal is monthly income and participates in online communities. He does not seek high returns and is not usually familiar with investments such as hedge funds. | monthly income in trade and communities |
| 1 | He knows fixed income well, but does not seek monthly income as a goal or high return. Many do not interact. | fixed income yes, but no monthly income in trade |
| 2 | Knows fixed income well and aims to build a monthly income. It was not brought to daytrade by influencers. Many do not interact. | fixed income and monthly income in trade |
| 3 | Know fixed income and cryptocurrencies and participate in online communities. High return and monthly income are possible goals. | treasure, satoshis and communities |
| 4 | It has been daytrading for over 24 months. He is not interested in fixed income and multimarket, he does not seek high returns. In general, it does not interact. | experienced disinterested |
| 5 | Seeks a career in the financial market and participates in communities. Monthly income is not objective. | career in the market |
| 6 | He discovered daytrade through an influencer. Seeks high return. Follows the market but does not interact with other daytraders. Eventually, he seeks monthly income and does not usually know multimarket. | influencers and high return |
| 7 | It has been daytrading for over 24 months. You know fixed income well, but you want high return. Most do not interact online with other traders or communities. | high return safely |
| 8 | He is a novice and answers "no" to learning about investments. Does not participate in communities, did not experience daytrade through influencers. | newbie |
| 9 | He discovered daytrade through an influencer. Search monthly income. It does not interact, it only follows the movements. In general, he is a novice and does not know hedge funds, etc. | influencers and monthly income |
| 10 | Seeks high return and knows cryptocurrencies more than fixed income. Does not interact much and was not brought by influencer. | satoshis and high return, interacts less |
| 11 | Knows fixed income well and seeks a monthly income. He has a lot of experience in the market. Know other funds, such as multimarket. Usually know crypto. Does not seek high return and not much to interact. | connoisseur |
| 12 | Main interest is participating in online communities. Search monthly income. Interacts strongly with others. He is not interested in a career in the market and he is not interested in funds such as multimarket or crypto. | communities and monthly income |
| 13 | Seeks high return. May have some experience in the market. Can interact a little, but does not participate in communities. Not much interest in hedge funds or fixed income | high return without much knowledge |

### 3.3 GrapeShape visual cluster validation

We introduce a visual metaphor to better understand clustering performance of binary datasets. Several visualization concepts are present in this technique. We assume that for each cluster we can calculate a representative centroid vector, as is the case of our result from the k-means clustering algorithm. We used the iconic representation, that is, we mapped a vector of values onto a figure that represents the various axes of the vector and changes proportionally to the values associated with these axes, in this case, the color of the grapes. Each grape represents a specific quiz answer and is a specific location within the icon, taking advantage of spatial memory. The color scale is designed to highlight the predictability of a "yes" value for the answer, mapped to purple, the predictability of a "no" value mapped to yellow, or the unpredictability of the answer value, mapped to "green".

One of the aspects that must be observed is the diversity of the bunches, that is, the bunches are quite different from each other. This shows that there is little overlap between the clusters.

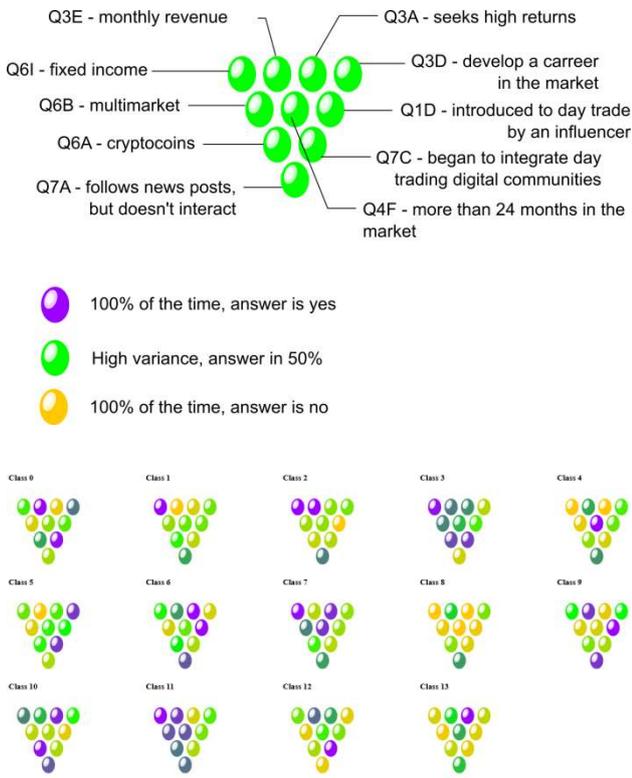

Figure 1. Iconic representation of customer groups obtained from questionnaire responses. The metaphor of bunches of grapes is used to show the diversity of the clusters and a color scale to inform the predictability of each response within the cluster members.

### 3.4 Visualization of relations between questions

In order to research relationships between questionnaire responses, we created a new graphic representation capable of expressing the variation in proportions and visually signaling the statistical significance of a hypothesis. From this analysis, we came up with the idea of using the conversion rate to perform the search for rules of interest. Pairs of variables were filtered in order to reduce the number of relationships to be analyzed. The final selection was done manually, producing a table of relevant relationships between questionnaire responses. This information can be consulted more easily if you have a specific objective and all selected ones have undergone a hypothesis test of proportions.

Let us explain through an example. Consider the statement: "Those who have a degree or a postgraduate degree in areas related to finance respond 10.8% more often that they were introduced to day trading by a brokerage firm than respondents to the questionnaire in general".

We are relating the answer "Q1C - presented by brokerage firm" with the answer "Q5E - graduation or postgraduate degree in the area". In the HalfPie chart below, in orange is the sector with angle proportional to the number of people who answered "yes" to "Q1C - presented by brokerage house", while in blue is the proportion of those who answered "no". There is a black line marking the division of sectors and whose angle corresponds to the proportion of this variable.

Now, instead of the totality of correspondents, we only consider people who answered "yes" to "Q5E - graduation or postgraduate degree in the area". In this case, the black line would be shifted to where the turquoise line is, showing that there is an increase in the proportion of Q1C when the Q5E response is positive.

See also that the magenta line represents the proportion of Q1C among those who answer "no" to Q5E. In addition, the turquoise line and magenta line have a border represented by a semi-transparent sector. This margin corresponds to the z-score confidence interval for two-sided p-value of 0.05. Thus, the graph also allows us to perform a hypothesis test visually. If the black line is outside the margin, the statement is statistically significant by the test of proportions. For example, The claim made that true Q5E increases the proportion of Q1C is valid and statistically significant. The statement that negative Q5E reduces the proportion of Q1C is not considered statistically significant since the black line is under the magenta margin.

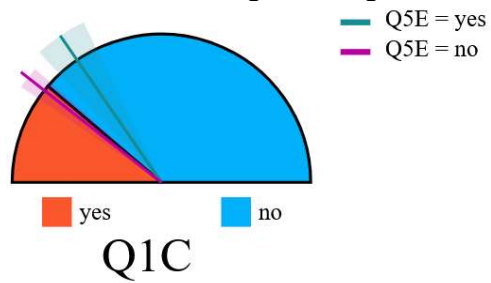

Figure 4. Example of a semi-pie chart to assess the relationship between two survey responses.

The graph is visually rich as it shows the proportions in the general and particular cases and these margins that allow the hypothesis to be confirmed. Thus, it is useful for quantitatively explaining the relationship and for detecting more important and pronounced relationships and quickly validating them by visual inspection.

In addition to the visual analysis, we can evaluate all pairs of responses by calculating the significance through the z-score and respective p-value [11]. Thus, we only filter relationships with p-value less than 0.05.

We also seek to omit relationships between responses to the same question. For example, we are not looking for relationships like "those who invest in fixed income also invest in direct treasury".

Even so, the number of relationships is large for the analysis, so that other information mining strategies need to be used. In this way, we define a metric to select the most interesting relationships from our point of view.

Let us consider another example, the statement: If the answer is positive for "Q3B - aim to compete with friends", then the frequency of "Q2G - podcast information source" is 19.8% higher.

When we look at the proportion only in the set of customers who answered yes to Q5E, the proportion of Q1C increases to where the green line is.

If the answer is positive for "Q3B - objective to compete with friends", then the frequency of "Q2G - podcast information source" is 19.8% higher.

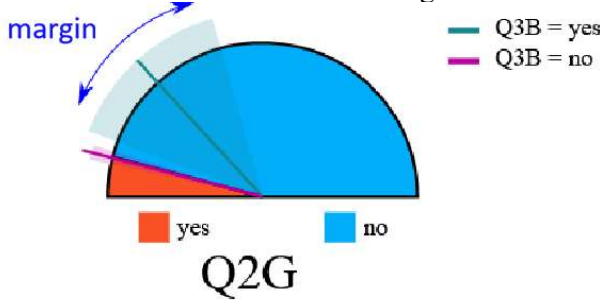

Figure 5. Example of a semi-pie chart to evaluate the relationship between using a podcast as a source of information and competing with friends as a trade objective.

Let's define a metric that we'll call "signed conversion rate" and that strongly represents the type of relationship between variables that we're looking for. If answer A is possibly influenced by answer B, we define the conversion rate as:

$$(2)$$

expression used when $P(A|B) > P(A)$

The conversion rate corresponds to the change in the proportion divided by the opening of the blue sector. That is, of the percentage of elements that answer no in the general case, what proportion of them would be "converted" with the inclusion of new information or condition. This value ranges from 0 to 100%.

When $P(A) > P(A|B)$, let's generalize and define the signed conversion rate. This time using the proportion of the orange sector as the denominator:

$$(3)$$

expression used when $P(A)>P(A|B)$ with a negative number as result. The value varies from -100% to 0.

Having this metric in hand, we can order the relationships, in order to highlight the most extreme ones. The margin indicates a confidence interval for a proportion hypothesis test (z-score) with two-tailed p-value of 5%.

Those who work in the market (support: 187 respondents) most often answer that they have more than 24 months of experience in day trading (with a conversion rate of 46.8%)

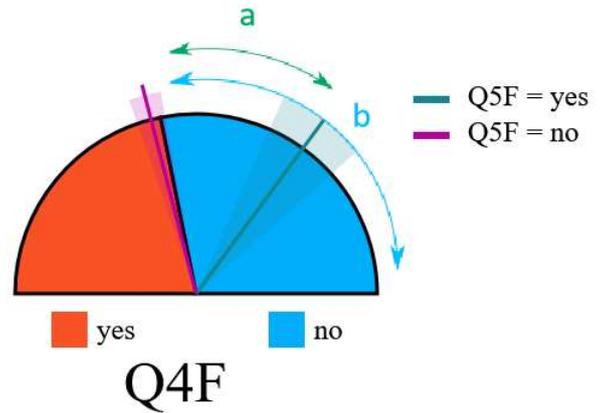

Figure 6. The positive response to "Q5F - work in the market" strongly influences the proportion of positive responses to "Q4F - more than 24 months of experience"

The a/b ratio is called the conversion rate and has been used as a criterion for selecting the rules that relate the answers.

Finally, let's give an example of a positive response causing a reduction in the ratio: Those who answer "Q4F - more than 24 months of experience" (support: 1340) respond less frequently that they use Youtube as a source of information (Q2L) with a negative conversion rate -5.2%.

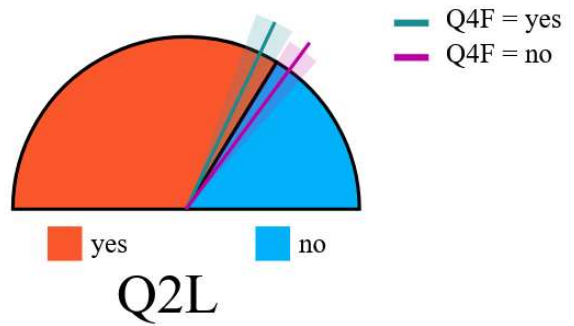

Figure 7. Example of negative result influence: Those with more experience (support 1340) respond less frequently that they use Youtube as a source of information (-5.2% negative conversion rate)

### 3.5 Automatic mining of relations

Criteria for selecting rule membership include:
1. P-value < 5%
2. Conversion rate (higher then a threshold)
3. Support (highr than a threshold)
4. Answers for questions in different groups
5. Non-obvious relationships

Next, we present some highlights found in the search for rules of influence between variables. Remembering that all pass the test of statistical significance of proportion. Example of obvious rule: "If source of news is friends, the chance of having been introduced to day trade by friends is 38.2% times higher and happened 299 times".

Selected insights extracted from the data:

1. Higher academic levels raise the chance of being introduced by a "corretora" (10.8%, 395).
2. Works in market increases the chance of being introduced to day trade by a bank (10.2%, 187)
3. Those who publish content in the internet are 23.1% more frequently users of instagram as a source of news (202).
4. Facebook users were 7.7% times more frequently introduce to day trade by advertisements. (34)
5. If objective is high returns, 21.5% less chance of not following news. (1293)
6. If objective is to know the market, 26.5% times more uses news sites. (491)
7. If objective is to interact socially, 34.6% more are telegram users (41). 41.3% more are Whatsapp users (41).
8. If introduced by influencer, 42.9% higher chance of using Youtube as a news source (1046).
9. If news source is Facebook, 15,1% higher the chance to seek high returns (226).
10. Content publishers respond 18.8% more often that objective is a carrer in the market.
11. If source is friends (25.4%, 299), "assessoria" (24.2%, 208), short course (19.7%, 474), discord (17.8%, 86), instagram (14.3%, 1053), the objective is more frequently a monthly income.
12. Source of information podcast 22.6% 255, digital communities 19.2% 1142, instagram 18.4% 1053, twitter 15.4% 217, friends 13.7% 299, Discord 13.5% 86, complete course 12.2% 1286, news sites 12.0% 1776 or books estudo: livros 11.5% 1121 is related to the desire for autonomous work style.
13. The objectives of competing with friends (16.6%, 34) and intellectual challenge (13.4%, 535) are related to more frequent book readers.
14. Having a complete course is related to "assessoria" (24.2%, 75), structured operations (23.3%, 649), digital communities (19.9%, 1142), Discord (17.9%, 86), publishing content (15.6%, 202).
15. Publishing content (19.1%, 202), using discord (18%, 86) or being introduced by an "assessoria" raises the chance of responding interest about cryptocoins.
16. Interest in multimarket funds is 31.6% more often by people working in the market (187).

### 3.6 Gender and age analysis

In order to further characterize the respondent sample, we describe and analyze by gender and age. The cluster analysis crossed with age and gender data was able to produce additional information about the profiles. It was possible to distinguish profiles with younger people and novices and profiles with more experienced people and, in a way, more novices are related to the profiles with the lowest performance in the financial analysis by clusters carried out in a separate chapter.

In Fig. 8, observing the first quartile, classes 4, 13 and 12 have older people. Observing by the third quartile, classes 5 and 6 contain the youngest. By the median, class 10 [crypto, high return, little interaction] stands out as the youngest. Class 12 is interesting because it has a certain seniority and at the same time has a lot of interest in digital communities. This profile characteristic goal is building a monthly income from day trading.

Analyzing the distribution of responses by age, we noticed, among other results, that those looking for a career in the market or already working in the market tend to be younger.

In the gender study (Fig. 9), we characterized some profiles with a greater female presence and identified that the monthly income objective is a possible characteristic of the female public, although we were unable to prove this statement statistically. In Fig. 10 we can notice that the black line in Q3E (goal: monthly income) is within the margin of the confidence interval for the woman category.

However, there is an overall gender-independent preference for this goal.

We also reveal other recurrent characteristics of the female audience, which are statistically significant: use of videos to educate themselves, having learned about day trading through friends, having an intermediate level of experience and knowing less about certain types of investments. Thus, a possible strategy to reach the female audience would be to communicate through friends and family or videos on the internet, educate in different investment modalities and teach how to build a monthly income in day trading.

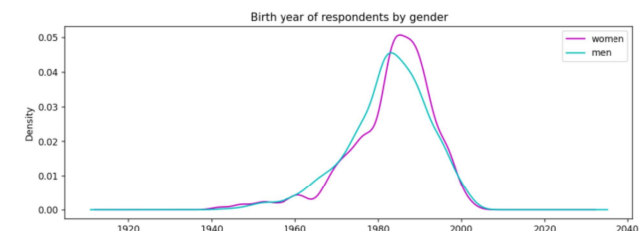

(a)

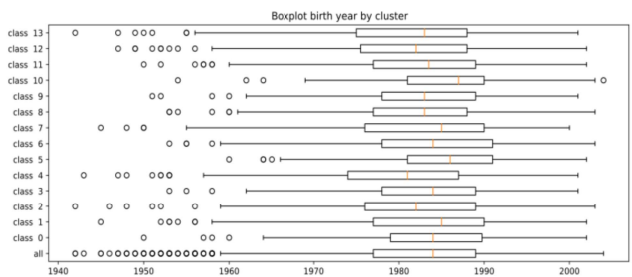

(b)

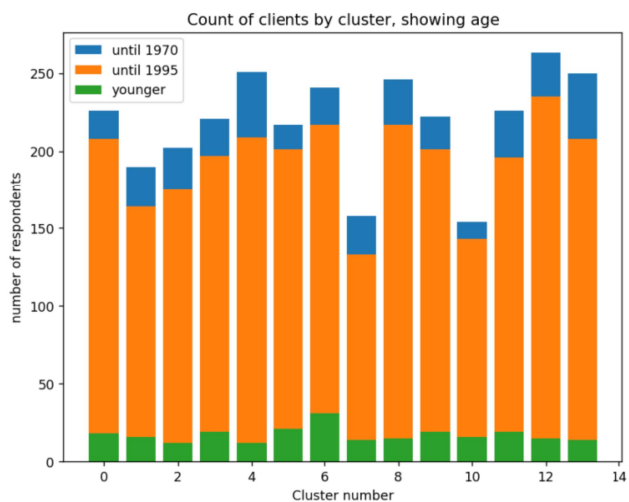

(c)

Figure 8. Respondent birth date distribution (a) per gender and (b) per cluster. (c) Cluster distribution including age.

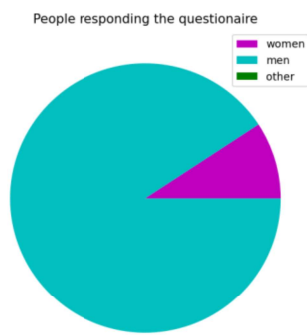

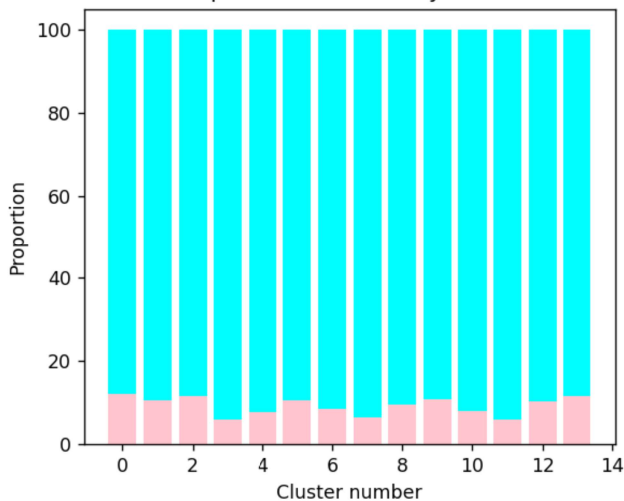

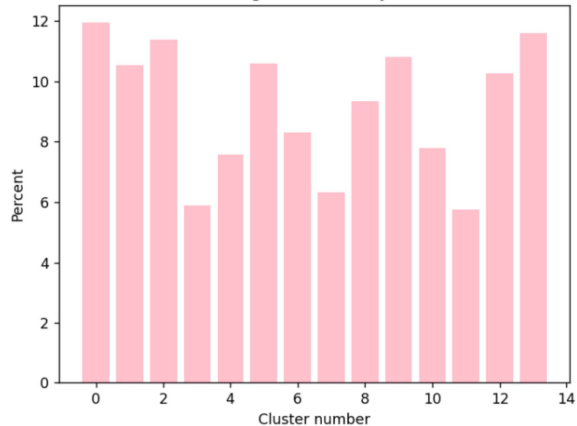

Figure 9. Gender proportion (a) overall (b, c) per cluster.

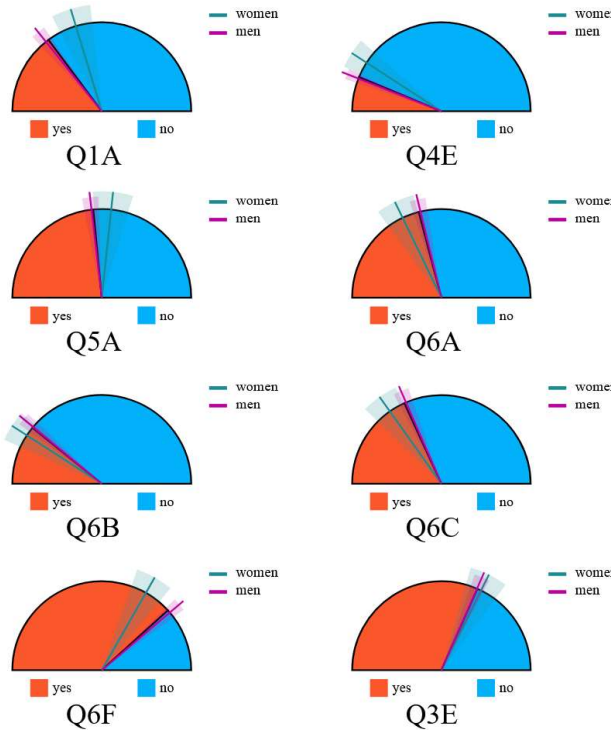

Figure 10. Analysis of women responses.

According to the HalfPie charts in Fig.10, the following items are most frequently answered by women in comparison to men:
- (Q1A) getting to know day trade through friends or family
- (Q4E) be 6 to 12 months in the day trade
- (Q5A) watched some videos and searched about the subject on the internet
- (Q6A, B, C, F) less knowledge of cryptocurrencies, multimarket, stock funds and Bovespa market.

## 3.7 Analysis against returns performance

By analyzing the operations and computing the returns in day trade (only considering operations that start and finish in the same day) we computed the performance of each respondent of the questionnaire. They were ranked and filtered by 5% best and 15% best. We then look at the cluster distributions (Fig. 11) of the general population in comparison to top performers. It is very clear that for the general population the cluster distribution is balanced, while there is high skewing when we consider only the top performers.

In addition, we look at how the top performers respond to the questionnaire, in comparison to the general set of respondents (Fig. 12).

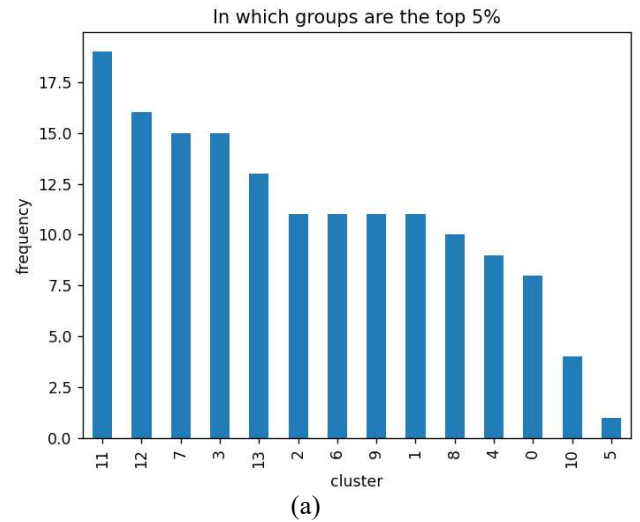

(a)

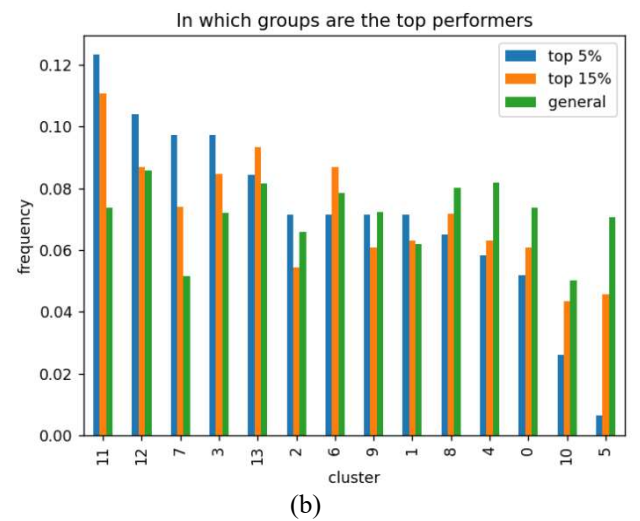

(b)

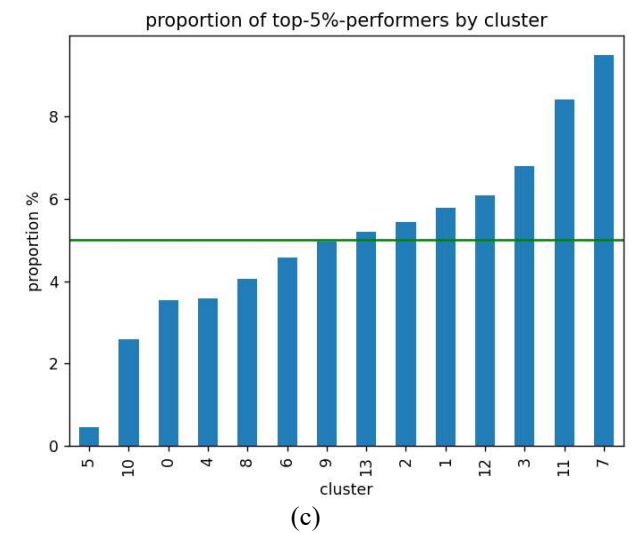

(c)

Figure 11. (a) The distribution of 5% top performers in each cluster, showing big unbalance, (b) same chart comparing 5%, 15% and general population, showing balance before selecting the top performances, (c) proportion of top performers in each cluster.

In short, the selected answers that are related to top performers are:
1. - Source of information: news site or not to follow news
2. - First contact: through a news site, or advisory, broker or friends
3. - Goal: monthly income, to know the market or high returns
4. - Experience: more than 24 months
5. - Study: short courses (less than 40 hr), already work in the market
6. - Investments: knowledge of all contribute, but non-financial, multimarket and fixed income stand out
7. - Digital interaction: not follow or interact

## 5. Discussion

Cluster analysis was effective in reducing the variance of groups of respondents and producing diversified profiles. Using the dimensionality reduction, a selection of responses more representative of the behaviors led to a gain in the internal metrics of the cluster. Many responses with low variance and consequent high predictability proved to be uninformative and did not contribute to the cluster analysis.

With the help of a new iconic representation for the clusters, it was possible to visualize the profiles formed. In addition, these profiles were described and received a stereotype to characterize their behavior.

Some analyzes were carried out with the crossing of information on date of birth and declared financial value applied, showing that the clusters can be interpreted and already beforehand, it was possible to highlight clusters 7, 11 and 13 with perspectives of good performance, and the clusters 5, 6, 8 and 10 as novice clusters. Thus, some clusters stood out as representative of good and bad performance behaviors. This was confirmed by analyzing the cluster distribution of the top performers.

We also analyzed how these top performing customers respond to the questionnaire questions in order to characterize their behaviors. In this way, we found that financial news sites are important. The best goals are monthly income, knowing the market or getting a high return. Experience counts and working in the market helps a lot. Knowledge about investments of any kind contributes to performance in futures and indices day trading, but non-financial investment, hedge funds and fixed income stood out.

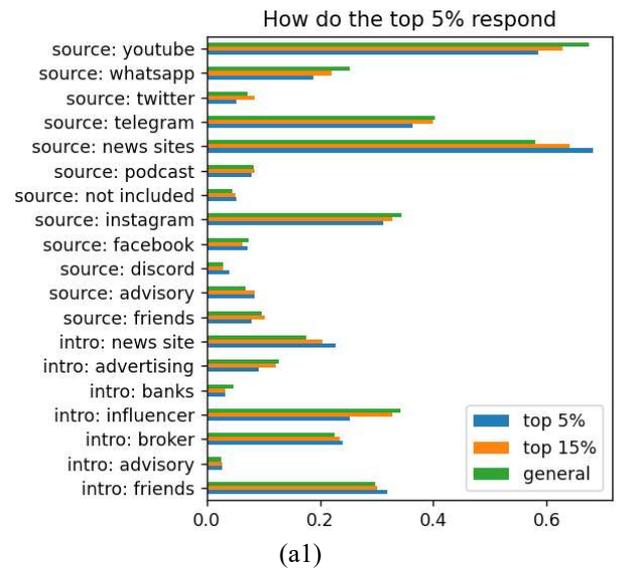

(a1)

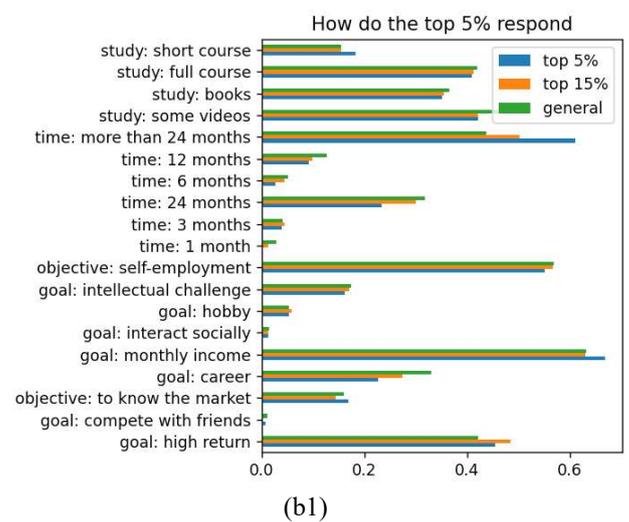

(b1)

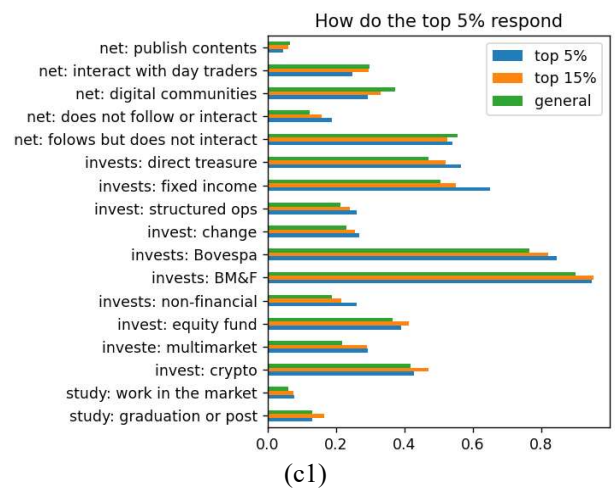

(c1)

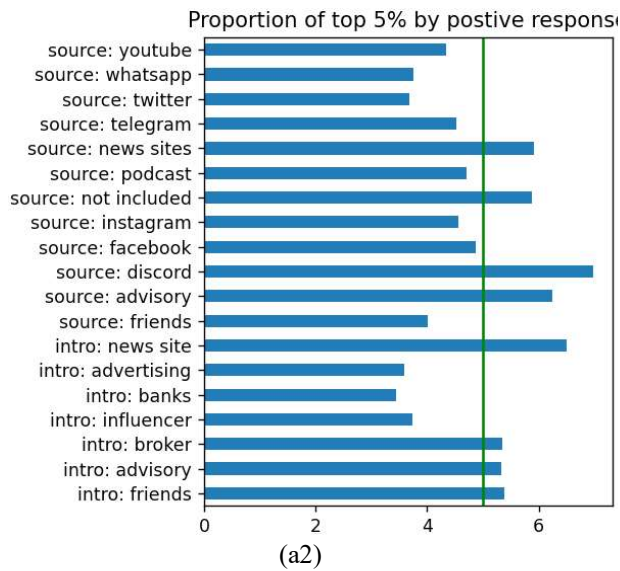

(a2)

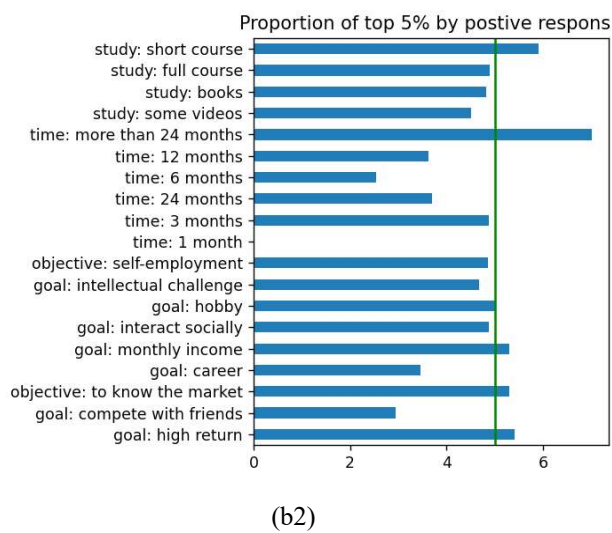

(b2)

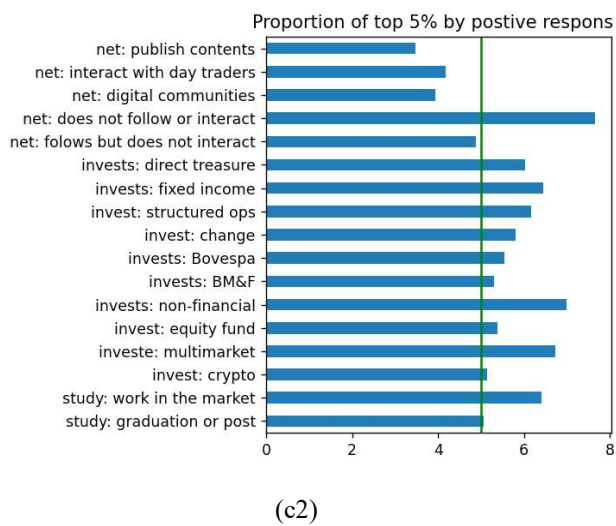

(c2)

Figure 12. (a1,b1,c1) frequency of "yes" responses per question for performance groups, (a2, b2, c2) proportion of "yes" responses per question by top performers.

These were unsurprising results, but we also discovered some more counter-intuitive information that deserve some interpretation. Top performers also answered that it is possible to perform well without following news and without interacting on the internet. Shorter courses also stood out as important source of information and techniques for a good performance. The most natural conclusion is that quality of the information matters, and selecting the source of information may help to evade the noise present on Internet, such as rumors and fake news.

It became clear that by separating the best performing customers, we were able to distinguish advantageous profiles and behaviors through the answers to the questionnaire. However, if we use the questionnaire responses to try to predict customer performance (whether it is in the top 15% or not) we fail terribly. Likewise, a regression analysis shows that there is little correlation between performance ranking and questionnaire responses.

The cluster analysis was effective in modeling behaviors, because, the distribution through the clusters was originally balanced, but, when we filtered by the 5% of customers with the best performance, there was a strong imbalance between the clusters. Thus, some clusters stood out as representative of good and bad performance behaviors.

Concerning the limitations of this work, one weak point is that the questionnaire was optional and the respondents were voluntary and self-appointed. However, no inconsistencies were perceived in the data and the rediscovery of expected relationships confirm that the questionnaire was seriously answered by league participants. A disadvantage of imposing the questionnaire is people not willing to collaborate, resulting in inconsistencies and noise.

## 6. Conclusion

We presented in this paper a methodology for data analysis using questionnaires with binary responses and exploratory data analysis methods: dimensionality reduction through clustering, respondent profile identification through clustering, visual cluster validation, automatic mining of rules concerning pairs of variables, visual validation of rules.

Regarding visualization techniques, we introduced the Grape Shape charts, a type of multiple chart displaying the characteristics of clusters of binary variable data. It allows the visual comparison among the clusters, but mainly the visualization of the internal cluster variance, giving an overview of the quality of clustering.

Another type of chart introduced is the HalfPie chart for a pair of binary variables. This chart allows the comparison of proportions of a variable with a confidence interval with regard to a second variable being positive or negative.

The process for mining rules concerning pairs of binary variables employed the "conversion rate" as a meaningful measure of proportion evolution of a variable with respect to another.

Finally, we presented a study case of the methodology applied to data from day trade investors that participated in a competition, revealing both expected and unexpected information regarding their behavior and background. In future work with this data, we intend to test different approaches to visualization such as the one in [12].

## Conflicts of Interest (Mandatory)

The authors declare no conflict of interest. The information here presented and shared is meant to improve knowledge on analysis methods of investor behaviour using questionnaires. It is reflex of the collected data as a case study. Information is provided in an aggregate fashion without identifying or exposing the individual investors.

## Author Contributions (Mandatory)

Carlos was responsible for the clustering analyzes, rule mining and innovative visualization methods. Paulo was responsible for processing financial information, connecting this to the questionnaire data and revising the paper. Andrei was responsible for providing information on the competition, evaluating the results and revising the paper. All of them and a few more members from XP participated in the design of the questionnaire.

## Acknowledgments

This work was supported by the ITA - XP agreement and the project was executed through Casemiro Montenegro Filho Foundation. Published with the permission of XP investimentos.